\documentclass[12pt]{iopart}

\input{epsf} 

\usepackage{graphicx}

\newcommand {\bc} {\begin{center}}
\newcommand {\ec} {\end{center}}
\newcommand {\bd}{\begin{displaymath}}
\newcommand {\ed}{\end{displaymath}}
 \newcommand {\be} {\begin{equation}}
\newcommand {\bea} {\begin{eqnarray} \nonumber }
\newcommand {\ee} {\end{equation}}
\newcommand {\eea} {\end{eqnarray}}
 \newcommand {\eps} {\epsilon}

\newcommand {\la} {\lambda}

\newcommand {\Si} {\Sigma}
 \newcommand {\al} {\alpha}

\newcommand {\lan} {\langle}
\newcommand {\ran} {\rangle}

\newcommand {\cD}  {{\cal D}}

\def\eps{\epsilon}
\def\al{\alpha}

\def\la{\langle}
\def\ra{\rangle}

 \def\(({\left(}
 \def\)){\right)}
\def\[[{\left[}
\def\]]{\right]}
\def\bi{\bibitem}

\newcommand {\for} {\ \ \ \mbox{for}\ \ }

\def \form#1 {eq. (\ref{#1}) }
\def \parziale#1#2  {{\partial {#1} \over \partial {#2}}}

\begin{document}
\title{Statistical Physics of Structural Glasses}
\author{Marc M\'ezard}
\address{Laboratoire de Physique Th\'eorique de l'Ecole
Normale Sup\'{e}rieure \footnote{UMR 8548:  Unit\'e Mixte du Centre National de la Recherche
Scientifique, et de
l'\'Ecole Normale Sup\'erieure}\\
24 rue
 Lhomond, F-75231 Paris Cedex 05, (France)\\
mezard@physique.ens.fr}
\author{Giorgio Parisi}
\address{Dipartimento di Fisica and Sezione INFN,\\
Universit\`a di Roma ``La Sapienza'',
Piazzale Aldo Moro 2,
I-00185 Rome (Italy)\\
giorgio.parisi@roma1.infn.it}

\date{\today}

\maketitle

\begin{abstract}
This paper gives an introduction and brief overview of
some of our recent work on the 
 equilibrium 
thermodynamics of glasses. We have focused onto  first principle computations in 
simple fragile glasses, starting
from the two body interatomic potential. A replica formulation translates 
this problem into
that of a gas of interacting molecules, each molecule being built of $m$ 
atoms, and having a 
gyration radius
(related  to the cage size) which vanishes at zero temperature. We use a 
small cage expansion, valid
at low temperatures, which allows to compute the cage size, the specific 
heat (which
follows the Dulong and Petit law), and  the configurational entropy. The
no-replica interpretation of the computations is also briefly described.
The results, particularly those concerning the Kauzmann
tempaerature and the configurational entropy, are compared to 
recent numerical simulations.
\end{abstract}

\section{Introduction}
While the experimental and phenomenological knowledge on
glasses has improved a lot in the last decades\cite{glass_revue}, 
the progress on a first
principle, statistical mechanical study of the glass phase has turned out
to be much more difficult. 

Take any elementary textbook on solid state physics.
It deals with a special class of solid state, the crystalline state, and 
usually
avoids to elaborate on the possibility of amorphous solid states. 
The reason  is very simple:
there is no theory of amorphous solid states. 
Schematically, the first elementary steps of the theory of crystals are 
the following.
 One computes the ground
state energy of all the crystalline structures. The small vibrations 
around these
structures are easily handled, either using the simple Einstein 
approximation
of independent atoms in harmonic traps, or computing the phonon
dispersion relations and going to the Debye
theory. Then one can study the one electron problem and compute the band 
structure.
The basic thermodynamic properties are already well reproduced by these 
elementary 
computations.
Anharmonic vibrations, electron-phonon and electron-electron interactions 
can then
be added to these basic building blocks.

Until very recently, none of the above computations, even in
the simplest-minded approximation,
could be done in the case of the glass state. The reason is obvious: all 
of them are
made possible in crystals by the existence of the symmetry group. The
absence of such a symmetry, which is a defining property of the glass 
state, 
forbids the use of all the solid state techniques. If one takes a 
snapshot of a glass
state, an instantaneous configuration of atoms, it looks more like a 
liquid
configuration. In fact the techniques which we shall use are often 
borrowed
from the theory of the liquid state. But while the liquid phase is 
ergodic (which
means that the probability distribution of positions is translationally 
invariant),
the glass phase is not. The problem is to describe a non-ergodic phase 
without a symmetry:
an amorphous solid state.

The work which we report on here has been elaborated during the last year 
and aims
at building the first steps of a first principle theory of glasses. The 
fact that this
is being made possible now is not  fortuitous, but rather results from
a conjunction of several sets of ideas, and the general progress of the 
last two
decades on the theory of amorphous systems. 

The oldest ingredients are the
phenomenological ideas, originating in the work of Kauzmann \cite{kauzmann}, and 
developed among others by
by Adam, Gibbs and Di-Marzio \cite{AdGibbs}, which identify the glass transition 
as a `bona fide'
thermodynamic transition blurred by some dynamical effects.
As we shall discuss below, in this scenario the transition is associated 
with an `entropy crisis', namely the
vanishing of the configurational entropy of the thermodynamically 
relevant glass
states.

 A very different, and more indirect, route, was the study of
spin glasses. These are also systems which freeze into amorphous solid 
states,
but one of their constitutive properties is very different from the 
glasses
we are interested in here: there exists in spin glasses some `quenched 
disorder':
 the exchange-interaction coupling constants between the spin
degrees of freedom are quenched (i.e. time independent on all 
experimental time scales)
 random variables\cite{rubber}.
 Anyhow, a few years after the replica symmetry breaking (RSB) 
solution of the mean field theory of spin glasses \cite{MPV}, it was 
realized 
that there exists   another category of mean-field spin
glasses where the transition is due to an entropy crisis \cite{REM}. These
are now called discontinuous spin glasses because their phase transition, 
although
it is of second order in the Ehrenfest sense, has a discontinuous order
parameter, as first shown in \cite{GrossMez}.
 Another name often found in the literature is 
 `one step RSB' spin glasses, because of the
special pattern of symmetry breaking involved in their solution. 
The simplest example of these is the Random Energy Model \cite{REM},
but many other such discontinuous spin glasses were found subsequently,
involving multispin interactions \cite{GrossMez,KiThWo,crisanti}.

The analogy
between the phase transition of discontinuous spin glasses and the 
thermodynamic
glass transition was first noticed by Kirkpatrick, Thirumalai and Wolynes
in a series of inspired papers of the mid-eighties \cite{KiThWo}. While some of 
the basic ideas
of the present development were around at that time, there still missed a
few crucial ingredients. On one hand one needed to get more confidence 
that
 this analogy was not just fortuitous. 
The big obstacle was the existence (in spin glasses) versus
the absence (in structural glasses)  of quenched disorder. The 
discovery  of discontinuous spin glasses without any 
quenched disorder
\cite{nodis1,nodis2,nodis3}
provided an important new piece of information: contrarily to what had 
been
believed for long, quenched disorder is not necessary for the existence of
a spin glass phase (but frustration is). A second confirmation came very 
recently from the developments on out of equilibrium dynamics of the 
glass phase.
Initiated by the exact solution of the dynamics in a discontinuous spin 
glass
by Cugliandolo and Kurchan \cite{cuku}, this line of research has made a lot 
of progress
in the last few years. It has become clear that, in realistic systems 
with short
range interactions, the pattern of replica symmetry breaking can be 
deduced
from the measurements of the violation of the fluctuation dissipation 
theorem \cite{fdr}.
Although these difficult measurements are not yet available, numerical
simulations performed on different types of glass forming
systems have provided an independent and spectacular confirmation of their
`one step rsb' structure \cite{gpglass,bk1,bk2} on the (short) time
scales which are accessible. The theory was then facing the big 
challenge:
understanding what this replica symmetry breaking could mean, in systems 
void of quenched disorder, in which there is thus no a priori
reason to introduce replicas. The recent progress has brought the answer 
to
this question and turned it into a computational method, allowing for
a first principle computation of the equilibrium thermodynamics of 
glasses \cite{MePa1,Me,MePa2,sferesoft,LJ,LJ2}.

In the context of glasses,
the words `equilibrium thermodynamics' call for some comments. First, 
it is
not obvious whether the glass phase is an equilibrium phase of matter. It 
might be
a metastable phase, reachable only by some fast enough quench, while the 
`true
equilibrium' phase would always be crystalline. The answer depends on the 
interaction potential. Numerically it is known that the frustration
induced by considering for instance binary mixtures of soft spheres
of different radii strongly inhibits crystallisation. But what is the
true equilibrium state is unknown,
and not very relevant. One can study crystals without having proven that 
they
are stable phases of matter (by the way, simply proving that the fcc-hcp 
is
the densest packing of hard spheres in  3 dimensions, a simple zero 
temperature
statement, has resisted the efforts of scientists for centuries
\cite{kepler}),
and one can study the properties of diamond, even though it is notoriously
unstable. The point is to have  reproducible properties, which is 
certainly
the case. Letting aside the crystal, a more interesting question is how
to reach equilibrium glass states. Experimentally nobody knows how to 
achieve
this. In a ferromagnet, one can reach an equilibrium state and eliminate 
domain walls
by using an external magnetic field. In a glass there is no such field 
conjugate
to the order parameter, and the fate is an out of equilibrium situation. 
The same
is true in spin glasses, and in fact in all kind of glass phases. Why 
study the
equilibrium thermodynamics then? The answer is twofold. First
principle computations are  certainly much easier as far as the 
equilibrium is concerned,
therefore it is natural to start with these in order to first get some
detailed understanding of the free energy landscape, which will be useful 
in the
more realistic dynamical studies. Secondly, we have strong indications, 
and some
general arguments, in favour of a close relationship between the 
equilibrium
properties and the observable out of equilibrium dynamical observations 
\cite{fdr}.
Let us also mention here the recent developments of some phenomenological
theory of the out of equilibrium theory of glasses \cite{theo}.

In this paper we shall introduce the main ideas of the recent elaboration 
of the
equilibrium theory of glasses. We shall not present the details which 
can be found in the literature. 
The general replica strategy can be found in \cite{remi,Me}. The
explicit  computations have been done first for soft spheres in
\cite{MePa1,MePa2}, and then generalized to binary mixtures of 
soft spheres \cite{sferesoft} or Lennard Jones particles \cite{LJ,LJ2}.

\section{Hypotheses on the glass phase}
The general framework of our approach is a familiar one in physics: we 
shall
start from a number of basic hypotheses on the glass phase, derive some 
quantitative
properties starting from these hypotheses, and then compare them with 
numerical,
and hopefully, in the future, experimental results. We work with a simple 
glass
former,  $N$ undistinguishable particles move in a volume $V$ of a 
d-dimensional space, and we take the thermodynamic limit $N,V \to \infty$ 
at fixed density 
$\rho=N/V$. The interaction potential is a two body one,
defined by a short range function $v(x)$ (for instance one may
consider a soft spheres system where $v(x)=1/x^{12}$).

\begin{figure}
\includegraphics[width=0.49\textwidth]{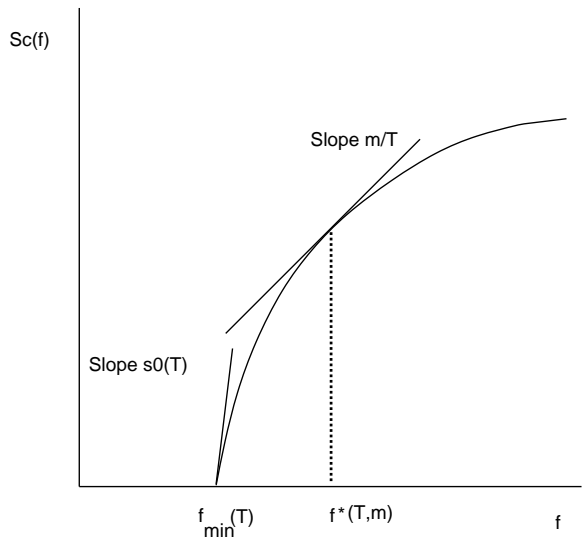}
\caption{Qualitative shape of the configurational entropy versus free energy.
The
whole curve depends on the temperature. The
saddle point which dominates the partition function,
for $m$ constrained replicas, is the point $f^*$ such
that the slope of the curve equals $m/T$ (for the usual unreplicated system,
$m=1$).
If the temperature is small enough the saddle point sticks
to the minimum $f=f_{min}$ and the system is in its glass phase. }
\label{sigma_qualit}
\end{figure}

Let us  introduce a free energy functional 
$F(\rho)$ 
which depends on the density $\rho(x)$ and on the temperature.  We 
suppose that at sufficiently low 
temperature this functional has many minima (i.e.  the number of minima 
goes to infinity with the 
number $N$ of particles).  Exactly at zero temperature these minima, 
labelled by
an index $\alpha$, coincide with the mimima of 
the potential energy as function of the coordinates of the particles.
A more detailed discussion of the valleys and their relationship to the
inherent structures \cite{inherent}  will be given in sect. \ref{entrop}. 
  To each valley we can associate a free energy $F_\al$ and a free energy 
density 
$f_\al= F_\al/N$. The number of free energy minima with 
free energy density  $f$ is supposed to be exponentially large:
\be
{\cal N}(f,T,N) \approx \exp(N\Sigma(f,T)),\label{CON}
\ee
where the function $\Sigma$ is called the complexity or the 
configurational entropy (it is the 
contribution to the entropy coming from the existence of an exponentially 
large number of locally 
stable configurations). This function is not defined in the regions 
$f>f_{max}(T)$ or $f<f_{min}(T)$, where ${\cal N}(f,T,N)=0$, it is convex 
and  
it is supposed to go to zero continuously
at $f_{min}(T)$,  as found in all existing models so far (see 
fig.\ref{sigma_qualit}). 
In 
the low temperature region the total free energy of the system, $\Phi$, 
can be well approximated by:
\be
 e^{-\beta N \Phi} \simeq \sum_\al e^{-\beta N f_\al(T)} =
\int_{f_{min}}^{f_{max}} df \ \exp\((N[ \Sigma(f,T)-\beta f]\)) \ ,
\label{SUM}
\ee
where $\beta=1/T$.
The minima which dominate the
sum are those with a free energy density $f^*$
 which minimizes the quantity $\Phi(f)=f-T\Sigma(f,T)$.
At large enough temperatures the saddle point is at $f>f_{min}(T)$. When 
one
decreases $T$ the saddle point free energy decreases.
The Kauzman temperature $T_K$ is that below which the saddle point sticks
to the minimum: $f^*=f_{min}(T)$. It is a genuine phase transition, the
`ideal glass transition'.

This scenario for the glass transition is precisely the one which is at 
work in discontinuous
spin glasses, and can be studied there in full details. The transition is 
of a rather special
type. It is of second order because the entropy and internal energy are 
continuous. 
When decreasing the temperature through $T_K$ there is a discontinuous 
decrease of
specific heat, as seen experimentally. On the other hand the order 
parameter is
discontinuous at the transition, as in first order transitions. To show 
this we have to provide
a definition of the order parameter in our framework of equilibrium 
statistical mechanics.
This is not totally trivial because of the lack of knowledge on the 
valleys themselves.
The best way is to introduce two identical copies of the system. We have 
one system
of undistinguishable `red' particles, 
interacting between themselves through $v(x)$, another
system of undistinguishable `blue' particles, interacting between 
themselves through $v(x)$, and we turn
on a small interaction between the blue and red particles, which is short 
range.
We take the thermodynamic limit first, and then send this red-blue 
coupling to zero.
If the position correlations between the red and blue particles disappear 
in
this double limit, the system is in a liquid phase, otherwise it is in a 
solid phase.
Clearly, the order parameter, which is the red-blue pair correlation
function, is discontinuous at the transition: there is no correlation
in the liquid phase, while in the solid phase one gets an oscillating 
pair correlation,
similar to that of a dense liquid, but with an extra peak at the origin.
 In some sense, in this framework, the role of the unknown
 conjugate field, needed in order to polarize the system into one state,
is played by the coupling to the second copy of the system.
The  small red-blue  coupling is here to insure that the two systems will 
fall into
the same glass state.

The above scenario, relating the glass transition to the vanishing
of the configurational entropy, is the main hypothesis of our work. 
Clearly
it is in agreement with the phenomenology
of the glass transition, and with the old ideas of Kauzman, Gibbs and 
Di-Marzio. It is also very interesting from the point of view of the
dynamical behaviour.

\begin{figure}
\includegraphics[width=0.49\textwidth]{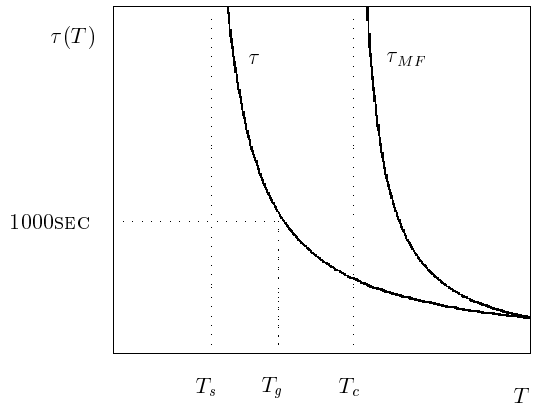}
\caption{Relaxation time versus temperature. The right hand curve is the prediction
of mode-coupling theory without any activated processes: it is a mean field prediction,
which is exact for instance in
 the discontinuous mean-field spin glasses. The left-hand curve is the observed
relaxation time in a glass. The mode coupling theory provides a quantitative prediction for the increase
of the relaxation time when decreasing temperature, at high enough temperature (well above the
mode coupling transition $T_c$). The departure from the mean field prediction at lower temperatures is 
usually attributed to 'hopping' or 'activated' processes, in which the system is trapped
for a long time in some valleys, but can eventually jump out of it. The ideal
glass transition, which  takes place at $T_s$, 
cannot be observed directly since the system 
becomes out of equilibrium on laboratory time scales at the `glass temperature' $T_g$.
Because of the special scenario of the static transition in mean field
spin glasses, due to some entropy crisis, the transition temperature
 $T_s$ should be identified with the Kauzman temperature $T_K$.  }
\label{figtau}
\end{figure}
 In discontinuous mean field spin glasses, the slowing down 
of the dynamics takes a very special form.
There exist a dynamical transition temperature $T_c>T_K$. When T decreases and gets near to $T_c$,
the correlation function relaxes with a characteristic two step forms: a
fast $\beta$ relaxation leading to a plateau takes place on a characteristic time
which does not grow, while the $\alpha$ relaxation from the
plateau takes place on a time scale which diverges when $T \to T_c$. This dynamic
transition is exactly described by the schematic mode coupling equations. The
existence of a dynamic relaxation at a temperature above the true thermodynamic one
is possible only in mean field, and the conjecture\cite{KiThWo}
 is that in a realistic system like a glass, 
the region between $T_K$ and $T_c$ will have instead a finite, but very rapidly 
increasing, relaxation time, as shown in fig. \ref{figtau}.

On this figure we see the existence of several temperature regimes:

-a relatively high temperature regime where mode coupling theory applies

- an intermediate region, extending from $T_k$ up to the temperature
above $T_c$ where mode coupling predictions start to be correct. This is the region
of activated processes, where one can identify some traps in phase space in which the system
stays for a long time, and then jumps.

-the low temperature, glass phase $T<T_K$. 

The dynamics of the glass is expected to show aging effects in the glass region, but also in
the intermediate region provided the laboratory time is smaller than the relaxation time.

Here we shall focus onto the equilibrium study of the low temperature phase. One main reason is
that the direct study of  out of equilibrium dynamics is more difficult, and that one might be able
to make progress by a careful analysis of the landscape \cite{angelani}. Another motivation
is to go into a more quantitative test of the basic scenario: while it agrees qualitatively
with several observations, as we just discussed, it should also be able to help
make more quantitative predictions.

 Our strategy
will be to start from this set of hypotheses and derive the quantitative
predictions which can be checked independently.
We shall be
able to compute for instance the configurational entropy versus free
energy within some well controlled approximations, and compare it to
the results of some numerical simulations.

\section{Replicas}
In order to cope with the degeneracy of glass states and the 
existence of a configurational entropy, a choice method is the
replica method. Initially replicas were introduced in order to
study systems with quenched disorder, in which one needs to compute
the disorder average of the logarithm of the partition function \cite{MPV}.
It took a few years to realize that a large amount of information is
encoded in the distribution of distances between replicas. This is
 true again in structural glasses. The simplest example was given above
when we explained the use of two replicas in order to define
the order parameter. A much more detailed information can be gained if
one studies in general a set of $m$ replicas, sometimes named `clones' in
this context, coupled through a small 
extensive attraction  which 
 will eventually go to zero \cite{remi,Me}.  In the glass phase, the 
attraction will force all $m$ systems
 to fall into the same glass state, so that
the  partition function is:
\be
Z_{m} = \sum_\al e^{-\beta Nm f_\al(T)}= \int_{f_{min}}^{f_{max}} df
 \ \exp\((N [ \Sigma(f,T)-m \beta f]\))
\label{zm}
\ee
In the limit where $m \to 1$ the corresponding partition function 
$Z_m$ is dominated by the correct saddle point $f^*$ for $T>T_K$. 
The interesting regime is when the temperature is  $T<T_K$, 
and the number $m$ is allowed to become smaller than one. The saddle 
point $f^*(m,T)$ in the expression (\ref{zm}) is the solution
of $\partial \Sigma(f,T) / \partial f=m/T$. Because of the
convexity of $\Sigma$ as function of $f$, the saddle point is
at $f>f_{min}(T)$ when $m$ is small enough, and it   sticks at 
$f^*=f_{min}(T)$
when $m$ becomes larger than a certain value $m=m^{*}(T)$,
a value which is smaller than one when $T<T_K$. The free energy
in the glass phase, $F(m=1,T)$, is equal to $ F(m^*(T),T)$. As the free 
energy
is continuous along the transition line $m=m^*(T)$, one can compute 
$F(m^*(T),T)$ from the region $m \le m^*(T)$, which is a region where the
replicated system is in the liquid phase. This is the clue to
the explicit computation of the free energy in the glass phase. 
It may sound a bit strange because one is tempted to think of $m$ as an 
integer
number. However the computation is much clearer if one sees $m$ as 
a real parameter in (\ref{zm}). As one considers low temperatures $T<T_K$ 
the
$m$ coupled replicas fall into the same glass state and thus they build
some molecules of $m$ atoms, each molecule being built from one atom of 
each 
'colour'. Now  the interaction strength of one such molecule with another 
one
is basically  rescaled by a factor $m$ (this 
statement becomes  exact in the limit of zero temperature
where the molecules become point like). If $m$ is small enough this 
interaction is small
 and the system of molecules is liquid. When $m$ increases, the molecular 
fluid
freezes into a glass state at the value $m=m^*(T)$.
So our method requires to estimate the 
 replicated free energy, 
$
F(m,T)=-{\log(Z_m) /( \beta m N )}
$,
 in a molecular
liquid phase, where the molecules consist of $m$ atoms and
$m$ is smaller than one. For $T<T_K$, $F(m,T)$ is maximum at
the value of $m=m^{*}$ smaller than one,
while for $T>T_K$ the maximum is reached at a  value $m^*$ is larger than one.
  The knowledge of $F_m$ as a 
function
of $m$ allows to reconstruct the configurational entropy
function $Sc(f)$ at a given temperature $T$
through a Legendre transform, using the parametric representation (easily
deduced from a saddle point 
evaluation of (\ref{zm})):
\be
f={\partial \[[m F(m,T)\]] \over \partial m} \ \ \ ; \ \  \Sigma(f)={m^2 \over T}
{\partial F(m,T)\over \partial m} \ .
\label{legend}
\ee

 The Kauzmann temperature ('ideal 
glass
temperature') is the one such that $m^*(T_K)=1$. For $T<T_K$ the equilibrium
configurational entropy vanishes. Above $T_K$ one obtains the equilibrium
configurational entropy $\Sigma(T)$ by solving (\ref{legend}) at $m=1$. 

More explicitly, one must thus 
introduce $m$ clones of 
each particle, with positions $x_i^a, a\in{1,...,m}$.
The replicated partition function is:
\bea
Z_m={1 \over N!^m} \int \prod_{i=1}^N \prod_{a=1}^m d x_i^a \ 
 \exp\((-\beta \sum_{1 \le i < j \le N} \ \sum_{a=1}^m v(x_i^a-x_j^a)
\right. \\ \left.
-\beta \epsilon \sum_{i,j=1}^N \ \sum_{1 \le a < b \le m}
w(x_i^a-x_j^b) \)) \ ,
\label{Z1}
\eea 
where $v$ is the original interparticle potential and $w$ is an
  attractive potential.
 This attractive potential must be of short range
(the range should be less than the
typical interparticle distance in the solid phase),
 but its precise form is irrelevant. Assuming
that $w$ is equal to $-1$ at very small distances,
and zero at large distances (notice
that the scale of the inter-replica interaction is fixed by the 
 parameter $\epsilon$), the coupling $w$ can be used to define an overlap between two 
configurations,
in a way similar to the crucial concept of overlaps in spin glasses. 
Taking
two configurations $x_i$ and $y_i$ of the $N$ particles, one defines the 
overlap 
between the configurations as
$q(x,y) = -1/N \sum_{i,k=1,N} w(x_{i}-y_{k})$, or the distance as $1-q$. 
The replicated
partition function with $m$ clones is thus (in more compact notations 
where 
$dx=\prod_{i=1}^N dx_i/N!$ and $H(x)\equiv \sum_{i<j}v(x_i-x_j)$ is 
the total energy of the system):
\be
Z_m=\int \prod_a dx^a \exp\(( -\beta \sum_{a}H(x_{a})+ \beta \eps N 
\sum_{a,b}q(x_{a},x_b)\)) \ .
\label{zm_ov}
\ee
This can be defined also for non integer $m$  using an analytic 
continuation
(if our hypothesis of the glass transition being of the same nature as 
the one
step rsb in spin glasses is correct, there is no replica symmetry 
breaking between
the clones\cite{remi,Me}, and the continuation is straightforward). 
Alternatively, one can
define it through the formula
\be
Z_m \propto \int d\mu(\phi) Z(\phi)^m
\ee
where $\phi$ is a quenched random potential defined in the full space,
which has a Gaussian distribution with moments:
\be 
\int d\mu(\phi)=1 \ \ ,\ \int d\mu(\phi) \  \phi(x)=0
 \ \ ,\ \int d\mu(\phi)\  \phi(x) \phi(y) =c^t- w(x-y)  \ ,
\ee
and $Z(\phi)$ is the partition function of one system in the external
potential $\phi$:
\be
Z(\phi)= \int dx \exp\((-\beta H(x) -\sqrt{\beta \epsilon} \sum_{i=1}^N\phi(x_{i})\)) \ .
\ee

\section{The molecular liquid}
The explicit computation of $Z_m$ in the regime $m<m^*(T)$ 
is a complicated problem of dense molecular 
liquids,
which requires some approximate treatments. Several types
of approximations have been developed recently, leading
to fully consistent results.
Focusing onto the low temperature regime, where the molecules  have a
small radius, it is natural to write the partition function in terms
of the center of mass and relative
coordinates $\{ r_i, u_i^a \}$, with $x_i^a=r_i+u_i^a$ and $\sum_a u_i^a=0$, 
and to
expand the interaction in powers of the relative displacements $u$. After
a proper renumbering of the particles, so that particles in the
same molecule have the same $i$ index, one gets:
\bea
Z_m &= &{1 \over N!} \int dr \prod _{a=1}^m du^a \prod_{i=1}^N 
\((m^3 \delta(\sum_{a=1}^m u_i^a) \)) 
 \exp\((-\beta \sum_{i<j,a} \[[v(r_i-r_j) \right. \right. \\
&& \left. \left. +\sum_{p=2}^\infty 
(u_i^a-u_j^a)^p
{v^{(p)}(r_i-r_j) \over p!}\]]
-{\eps \over 4} \sum_{i,a,b} (u_i^a-u_i^b)^2  \))
\ .
\label{zexpanded}
\eea
The last term is the small inter-replica coupling ($\eps$ will be
sent to zero in the end), which we have approximated for
convenience by its quadratic approximation. 
The expression (\ref{zexpanded}) can be expanded, at low 
temperatures,
in the following ways:
\begin{itemize}
\item
`Harmonic resummation': One keeps only the $p=2$ term. The action is 
quadratic
in $u$, and after performing the exact $u$ integral one obtains
an  effective interaction for the center of mass degrees of freedom,
which we shall detail below. The parameter $m$ appears as a coupling 
constant,
the analytic continuation in $m$ is thus trivial, 
and the whole problem reduces to treating the liquid
of center of masses, interacting through the effective interaction.
\item
`Small cage expansion': One expands the exponential in powers of the 
relative 
variables $u$, keeping only the $\eps$ term in the exponent. Again, the
$u$ integrals can be done exactly to each order
of the approximation. In this way one generates
an expansion of the free energy in powers of $1/\eps$. This function can 
be
Legendre transformed with respect to $\eps$, leading to a generalized free
energy expressed as a series in terms of the `cage radius', $ 
A=2/(3m(m-1)) \sum_{a,b} 
<(u_i^a-u_i^b)^2>$. Notice that the $1/\eps$ expansion is just  an 
intermediate step
in order to generate the small $A$ expansion of the potential (the same 
can be done
for instance when computing the Gibbs potential of an Ising model in 
terms of the
magnetization $M$ at low temperatures: even if one is interested in the 
zero magnetic
field case, one can introduce the field as an intermediate device and 
first
expand in powers of $\exp(-\beta h)$, before turning
the result into
an expansion in $1-M$).
\end{itemize}
The two methods are complementary. They both lead to the study of a 
liquid of
center of mass positions. The
small cage expansion is simpler because the result is expressed in terms 
of various
correlation functions
of the pure liquid of center of masses at the effective
temperature $T/m$, which can be  handled using traditional
liquid state techniques. On the other hand the
leading ($p=2$) term at low temperatures is not treated exactly.
In the harmonic resummation scheme the interaction
potential of the center of masses is modified: one gets
\be
Z_m= Z_m^0  \int dr
\exp\((-\beta m H(r) -{m-1 \over 2} Tr \log  M \)) 
\label{Zharmo}
\ee
where $ Z_m^0 ={m^{Nd/2} \sqrt{2 \pi T}^{N d (m-1)} / N!}$, and
the matrix $M$, of dimension $dN \times dN$, is given by:
\be
M_{(i \mu) (j \nu)}= {\partial^2 H(r) \over \partial r_i^\mu \partial r_j ^\nu}
= \delta_{ij} \sum_k v_{\mu\nu}(r_i-r_k)-  
v_{\mu\nu}(r_i-r_j)
\ee
and $v_{\mu\nu}(r) =\partial^2 v /\partial r_\mu \partial r_\nu$ 
(the indices $\mu$ and $\nu$ denote space directions). The effective 
interaction
contains the complicated `$Tr \log M$' piece
which is not a pair potential. Because of this term, in the whole glass 
phase where
one is interested in the $m<1$ regime, the partition function receives some 
contributions only from
those configurations $r_i$ such that all eigenvalues
of $M$ are positive: these are locally stable glass configurations. In order
to handle  this additional constraint, we used so far the following
 (rather crude) approximate  treatment, which consists of two steps. 
 First, a 'quenched approximation', which amounts to
 neglecting the feedback of
vibration modes onto the centers of masses, substitutes
$\la \exp\((-{m-1 \over 2} Tr \log  M \)) \ra$ by 
$ \exp\((-{m-1 \over 2} \la Tr \log  M \ra\)) $, where $\la . \ra$ 
is the Boltzmann expectation value at
the effective temperature $T/m$.
One is then left with the computation of the spectrum of $M$ in a liquid. 
This is an
interesting problem in itself. The treatment done in \cite{MePa1,MePa2} corresponds 
to keeping the
leading term in a high density limit. Further recent progress \cite{INM,MEPAZEE,CAGIAPA} 
should allow
for a better controlled approximation of the spectrum.

We shall not review here the details of these computations, which can be 
found
in \cite{MePa2} as far as the simple glass former with the
`soft sphere' $1/x^{12}$ potential is 
concerned, in \cite{sferesoft} for the mixtures of soft spheres and
in \cite{LJ,LJ2} for mixtures of Lennard-Jones 
particles.
Once one has derived an expression for the replicated free energy, one can
deduce from it the whole thermodynamics, as described above.
In all three cases, one finds an estimate of the Kauzman temperature which is in 
reasonable agreement with simulations, with a jump in specific
heat, from a liquid value at $T>T_K$ to the Dulong-Petit value
$C=3/2$ (we have included only positional
degrees of freedom) below $T_K$.
This is similar to the experimental result, where the glass
specific heat jumps down to the crystal value when one decreases the temperature
(Our approximations so far are similar to the Einstein approximation of
independent vibrations of atoms, in which case the contribution
of positional degrees of freedom to the crystal specific heat is $C=3/2$).
The parameter $m^*(T)$ and the cages sizes
are nearly linear with temperature in the whole glass phase. 
This means, in particular, that the effective temperature $T/ m$ is always 
close to $T_K$, so in our theoretical computation we need
only to evaluate the expectation values of observables in the liquid phase, 
at temperatures where the HNC approximation for the liquid still works quite well. 

A more detailed numerical checks of these analytical predictions involves
the measurement of the configurational entropy. We shall review these checks in
sect. \ref{entrop}, but we first wish to present some
alternative derivation of the low temperature results.

\section{Without replicas}
For those who do not appreciate the beauty and efficacy of the replica approach,
it may be useful to derive some of the above results without resorting to the
replica method \cite{pedago}. Specifically, we shall study the simplest case of
the zero temperature limit in the harmonic approximation through
 a direct approach, and reinterpret the above results.
At low temperatures, the critical value $m^*$ of the parameter $m$ goes to zero linearly with $T$.
We thus write $\gamma=\beta m$ and take the $T,m \to 0$ limit of (\ref{Zharmo})
at fixed $\gamma$. This gives:
\be
Z_m \simeq \(( {\gamma \over 2 \pi}\))^{Nd/2} \int_C dr \ 
 \sqrt{\det M(r)} \exp\((-\gamma H(r)\))\ ,
\label{Zm_lowT}
\ee
where $\int_C$ is restricted to configurations in which all eigenvalues
of $M$ are positive. 
A direct derivation of this formula, making all hypotheses
explicit, is the following. At zero temperature one 
is interested in configurations where every particle is in equilibrium:
$\forall i,\mu, \ {\partial H / \partial x_i^\mu} =0$. The number of such configurations
at energy $NE$,
\be
\mu(E)= \int dx \  | det M(x)|\  \delta(NE-H(x)) \prod_{i,\mu}
 \delta\(({\partial H \over \partial x_i^\mu}\)) \ ,
\ee
can be approximated at low enough energy,
where most extrema are minima \cite{CAGIAPA}, by the expression
\be
\nu(E)= \int_C dx \   det M(x) \  \delta(NE-H(x)) \prod_{i,\mu}
 \delta\(({\partial H \over \partial x_i^\mu}\)) \ .
\ee
Within this approximation $\nu(E)$ is related to the configurational entropy 
through $\nu(E)=\exp(N\Sigma(E))$,
and one can compute its Laplace transform:
\be
\zeta(\gamma) \equiv \int dE \  \nu(E) \exp\((-\gamma N E\))
= \int dE \exp\((N\[[\Sigma(E)-\gamma E\]]\)) \ .
\ee
Using an exponential representation of the ground state constraints,
this effective partition function is:
\be
\zeta(\gamma)= \(( { \gamma \over 2 \pi} \))^{Nd} 
\int dx  \prod_k d \lambda_k^\mu  \ det M(x) \  \exp\((-\gamma H(x)+
i \gamma \sum_{k,\mu} \lambda_k^\mu {\partial H \over \partial x_k^\mu} \)) 
\label{zeta}
\ee
One can change variables from $x_k$ to $y_k=x_k-i\lambda_k$. At low temperatures it is reasonable to assume that the only configurations which contribute
are those in the neighborhood of the minima. Expanding in powers of $\lambda$,
and neglecting anharmonic terms,
one writes:
\bea
H(x)&\simeq&H(y) +i \sum_{k,\mu} \lambda_k^\mu {\partial H(y) \over \partial y_k^\mu}
-{1 \over 2} \sum_{k,\mu,l,\nu} \lambda_k^\mu \lambda_l^\nu {\partial^2 H(y) \over \partial y_k^\mu
\partial y_l^\nu}\\
{\partial H(x) \over \partial x_k^\mu} &\simeq& {\partial H(y) \over \partial y_k^\mu}
+ \sum_{l,\nu}  \lambda_l^\nu {\partial^2 H(y) \over \partial y_k^\mu
\partial y_l^\nu} \ .
\eea
The $\lambda$ integral in (\ref{zeta}) is then quadratic, and one gets:
\be
\zeta(\gamma)= \(( { \gamma \over 2 \pi} \))^{Nd/2} 
\int_C dy\  \sqrt{ det M(y)} \  \exp(-\gamma H(y)) \ ,
\ee
a result identical to the low $T$ limit (\ref{Zm_lowT}) of the replica approach within the
harmonic approximation.

\section{Configurational entropy: theory and simulations}
\label{entrop}
The configurational entropy (sometimes called also complexity) is a key concept
 in the theory of 
glasses.
There is no difficulty of principle in defining a valley and its entropy in
the low temperature phase $T<T_K$. As we have seen, we can take a 
thermalized configuration as a reference system, add a small attraction 
to this configuration, and take the thermodynamic limit before the limit of 
a vanishing attraction. This procedure defines the restricted partition function
in the valley containing the reference configuration $y$,
and therefore the free energy of the valley. Computing $S_c(f,T)$ is thus in principle
doable, but it is still a formidable challenge to get equilibrated configurations $y$
in this temperature range.

On the other hand in the intermediate temperature regime $T_K<T<T_c$, the valleys and
the configurational entropy remain well defined in the mean field theory. The existence
of a decoupling of time scales
 points
to the possibility of defining metastable valleys in the whole region where 
activated ('hopping') processes are found. This region is particularly interesting,
both because of the rapid change of relaxation times,
and because part of this region can be studied experimentally  or numerically. 
It often happens that different authors use different 
definitions of the configurational entropy, which should be hopefully be 
equivalent at low temperature 
but behave rather differently at high temperatures.
Therefore
it seems to us appropriate to start
this section with a comparison of  the various definitions of configurational entropies
which have been introduced and studied so far.

If we consider the configurational entropy versus temperature, which is non-zero 
for $T>T_K$, in a first approximation we can distinguish three different types of 
definitions:
\begin{itemize}
    \item  A first definition is based on the presence of many minima of the Hamiltonian, i.e. 
    inherent structures.
     \item  A second definition is based on the fact that the phase space at 
sufficient low energy 
    may be decomposed in many disconnected region (let us call it the microcanonical one).
    \item  A third definition is based on the thermodynamics. One starts from the definition 
    \be
    S(T)=\Sigma(T)+S_{valley}(T)
    \ee
    where $S(T)$ is the total entropy and $S_{valley}$ is the entropy of the generic valley at 
    temperature $T$.  In this case the problem consists in finding a precise definition of $S_{valley}$.
\end{itemize}
In this paper we have used the third definition, however we think useful to recall the other two 
definitions in order to avoid possible misunderstanding.
\subsection{The inherent structure entropy}
  Given the  Hamiltonian $H(x)$ of a system with $N$ particles, we can 
consider the solution $x(t)$ of the equation
\be
{dx \over dt}=- {\partial H \over \partial x}
\ee
as function of the initial conditions $x(0)$. At large time $x(t)$ will go to one of the minima of the 
Hamiltonian, called an  inherent structure. 
We  label by $a$ each coherent structure and we  call $\cD_{a}$ the set of those 
configurations which for large times go to the coherent structure labeled by $a$.
The union of all the sets $\cD_{a}$ is the whole phase space.
The probability of finding the system at a temperature $T$ inside a given inherent structure is 
proportional to  
\be
P(a)=Z(a)/\sum_b Z(b) \ \ \  ; \ \ \  Z(a)\equiv \int_{x \in \cD_{a} }dx \exp (-\beta H(x)) \ .
\label{za}
\ee

The configurational entropy density, $\Si_{is}$, is defined by
\be
N \Si_{is}(T)= -\sum_{a} P(a)\ln(P(a)).
\label{Sis}
\ee
This definition makes sense at all temperatures. In the limit of large $T$ one finds
\be
\lim_{T\to\infty} \Si_{is}(T) =-\sum_{a} V(a)\ln(V(a)),
\ee
where $V(a)$ is proportional to the volume in phase space of the region $\cD_{a}$, normalized in 
such a way that $\sum_{a}V(a)=1$.
It is reasonable to expect that this inherent-structures configurational
 entropy starts to decrease when the temperature is decreased around $T=T_{c}$ and 
vanishes at $T=T_{K}$.

\subsection{Microcanonical entropy}
 We consider the hypersurface of constant energy density, 
$
H(x)=EN
$,
and decompose this energy surface in connected components which we label by $a$. The number of 
connected components clearly depends on $E$. 

Calling $V_{a}$ the normalized phase space volume of each connected component, we define the
microcanonical configurational entropy density as
\be
N \hat\Si_{m}(E)=-\sum_{a} V(a)\ln(V(a))
\ee
The microcanonical configurational entropy density as function of the temperature is naturally 
defined as
\be
 \Si_{m}(T) =\hat\Si_{m}(E(T))
 \ee
where $E(T)$ is the internal energy density as function of the temperature.
It is clear that at high energies the configuration space contains only one connected component and 
therefore 
\be
\lim_{T\to\infty} \Si_{m}(T)=0
\ee

The two configurational entropies introduced so far,
$\Si_{is}(T)$ and $\Si_m(T)$ certainly differ at 
high temperature and many hands must be waved in 
order to argue that both entropies behave in a similar way at low temperature and  vanish together at 
$T_{K}$.

\subsection{The thermodynamic configurational entropy}
As we have already stated  the thermodynamic configurational entropy can be defined by the 
relation
\be
\Sigma_{t}=S(T)-S_{valley}(T)
\label{Stdef}
\ee
The main difficulty is the precise definition of the valleys, and of  $S_{valley}(T)$,
in the regime $T>T_K$ where the system is still ergodic.  The basic idea \cite{SPEEDY} is 
to take a generic equilibrium configuration ($y$) at temperature $T$ and to define $S_{valley}(T)$ 
as the thermodynamic entropy of the system constrained to stay at a distance not too large 
 from the 
equilibrium configuration $y$.
If we impose a strong constraint (i.e. $x$ too near to $y$) the entropy will depend on the 
constraint, but the constraint cannot be taken vanishingly small
because the system is ergodic.

One may be worried that this method contains an unavoidable ambiguity. It turns out that
there exists a way to modify this method slightly in order to get rid of this ambiguity.
The modified method was introduced in \cite{pot} and called the potential method. Let
us summarize it here briefly.
Given two configurations $x$ and $y$ we define their overlap as before as
$
q(x,y) = -1/N \sum_{i,k=1,N} w(x_{i}-y_{k}),
$
where $w(x)=-1 \for x$ small, $w(x)=0 \for x$ larger than the typical interatomic distance.
Instead of adding a strict constraint we add an extra term to the Hamiltonian:
we define
\bea
\exp (-N \beta F(y,\eps))=\int dx \exp( -H(x)+ \beta  \eps N q(x,y)),\\
F(\eps)=\lan  F(y,\eps) \ran ,
\eea
where $\lan f(y) \ran$ denotes the average value of $f$ over equilibrium configurations $y$
thermalized at temperature $\beta^{-1}$.

\begin{figure}
\includegraphics[width=0.45\textwidth]{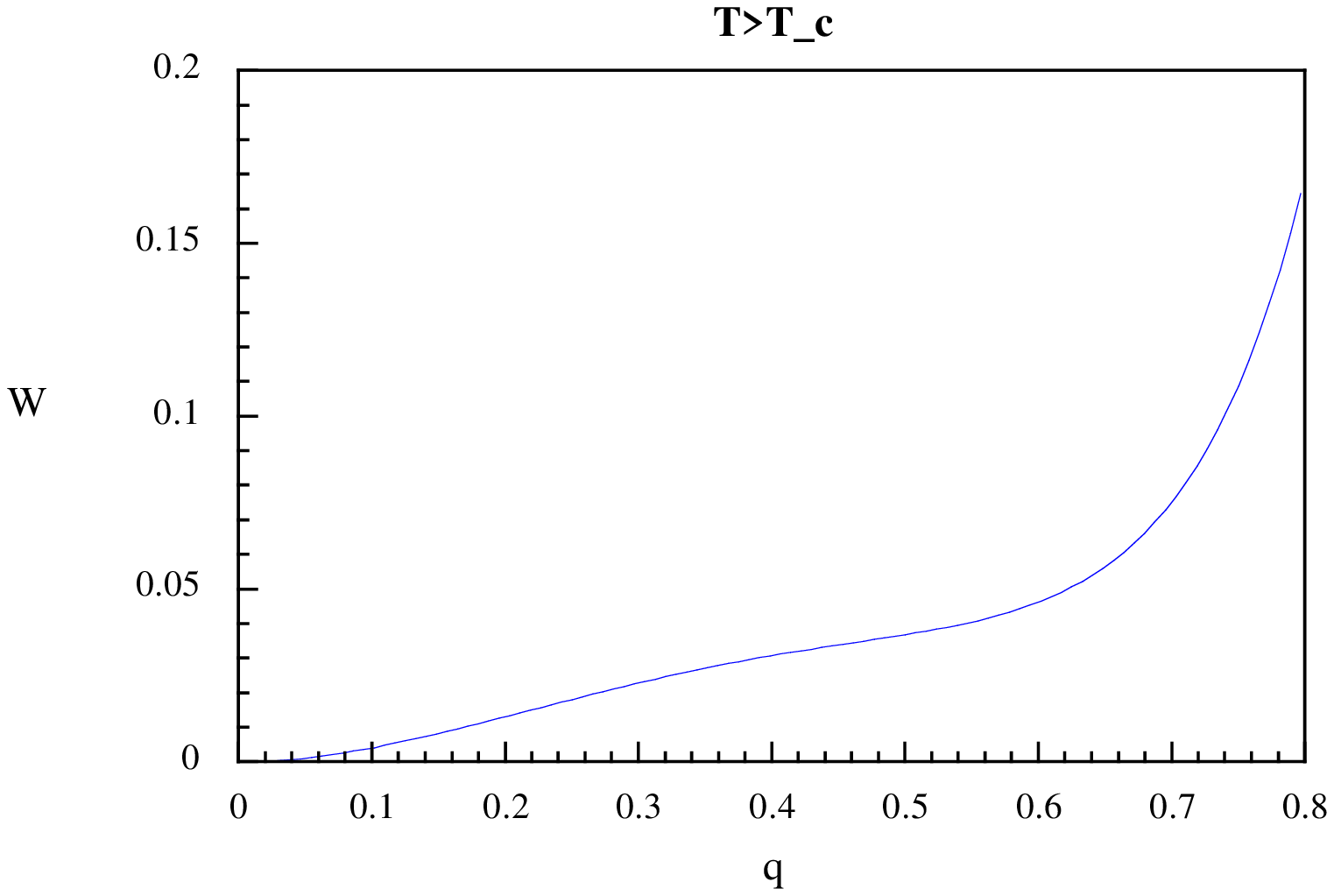}
\includegraphics[width=0.45\textwidth]{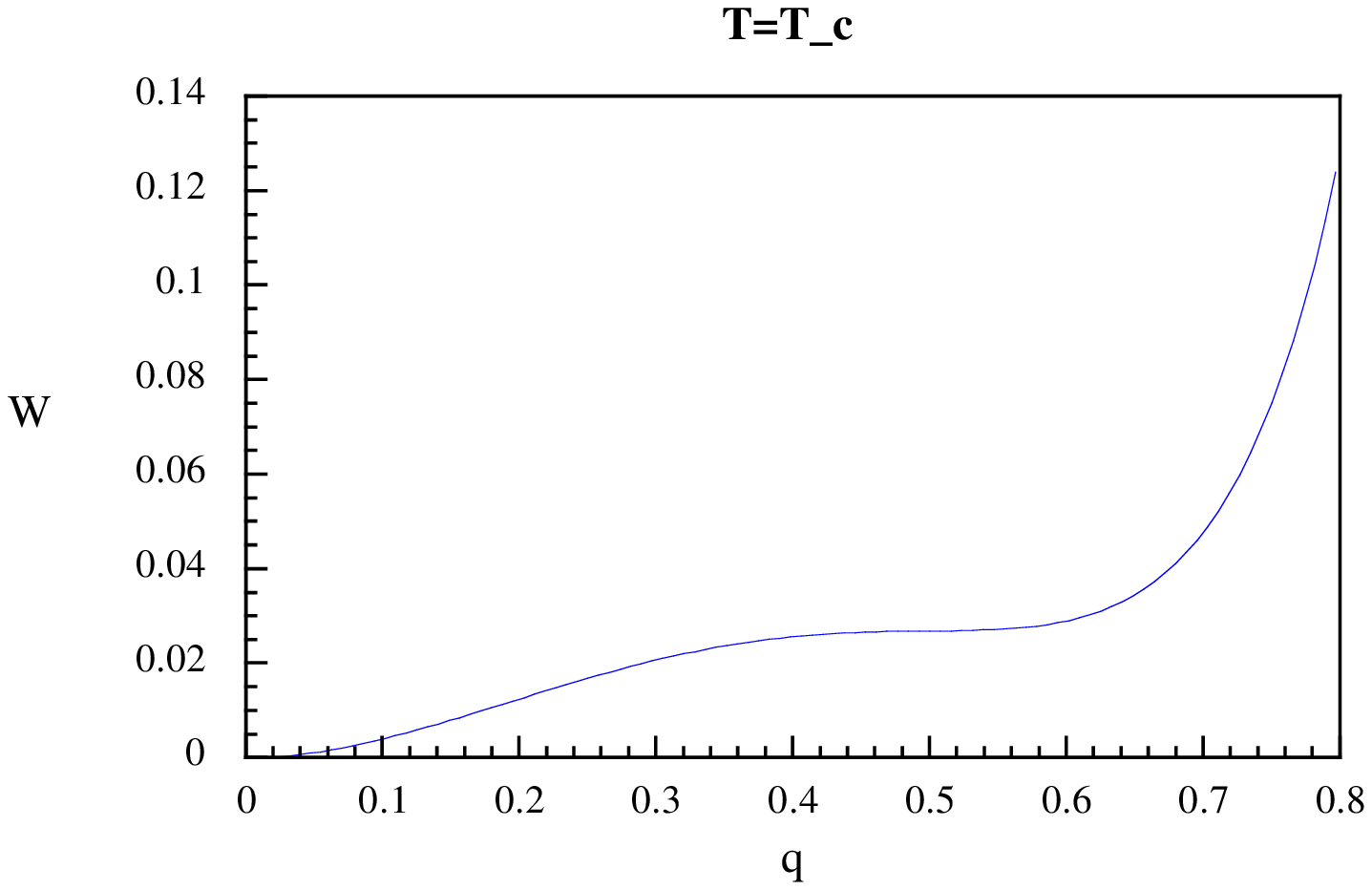}

\includegraphics[width=0.45\textwidth]{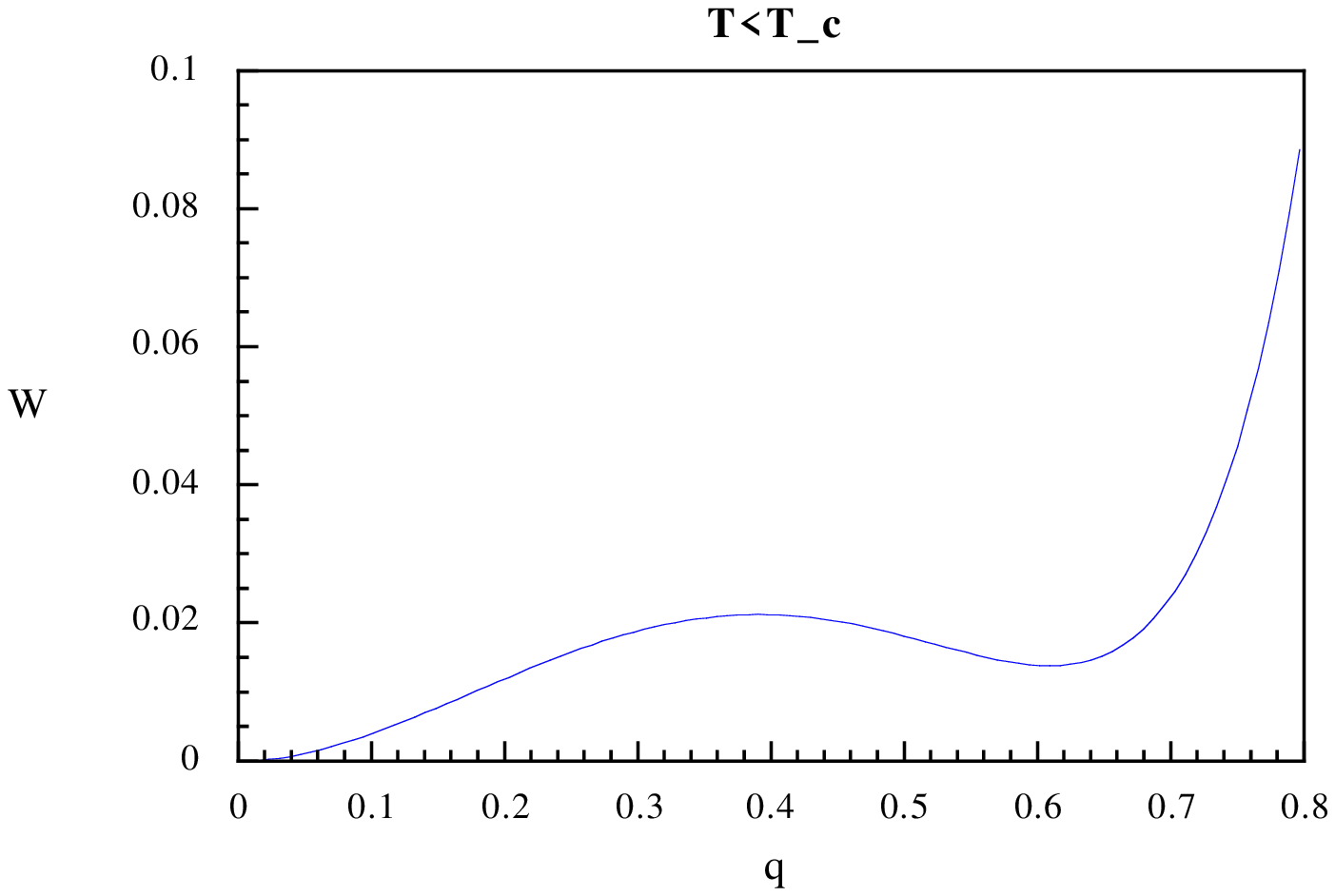}
\includegraphics[width=0.45\textwidth]{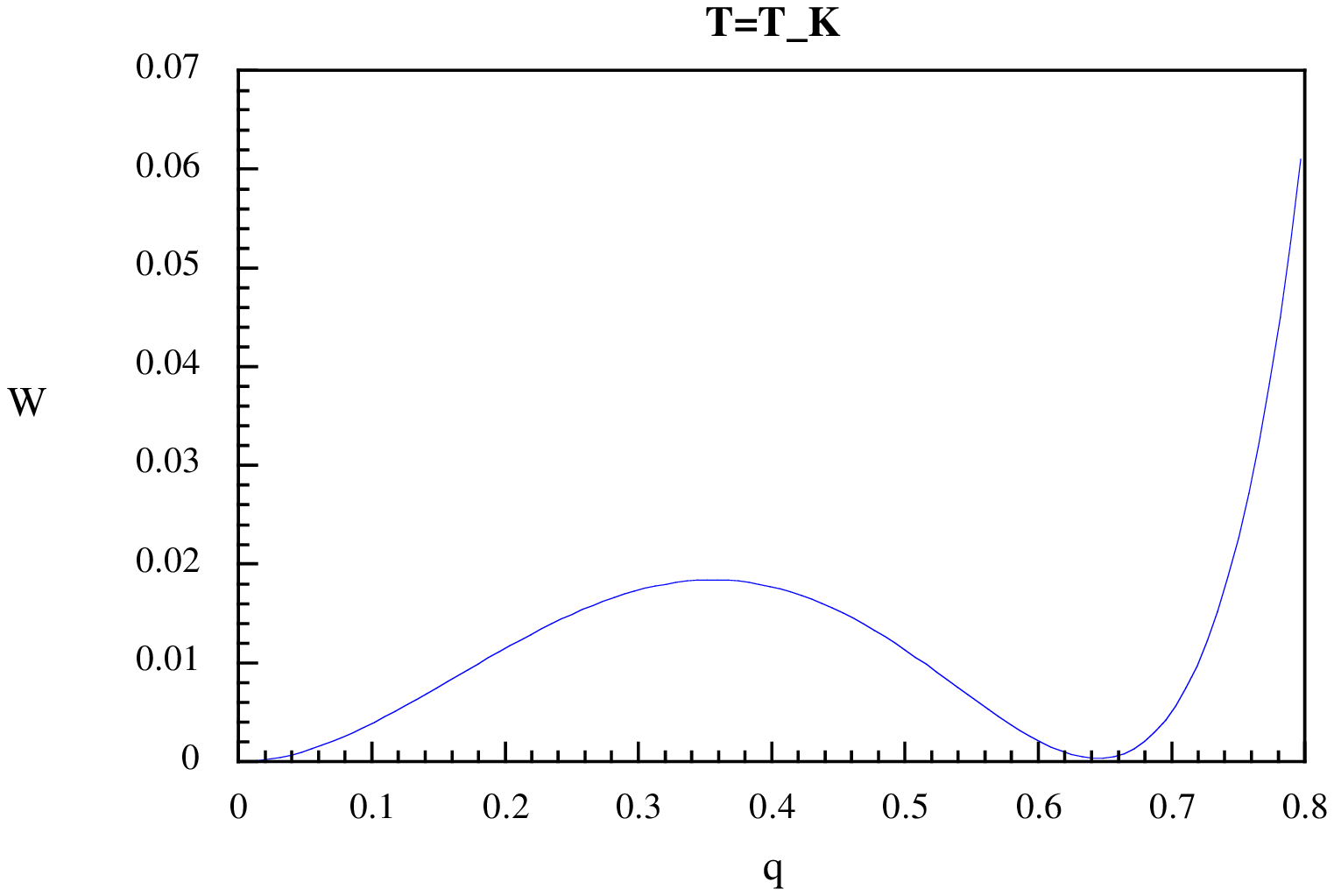}
\caption{Qualitative behaviour of the potential $W(q)$, in the four
 regions
$T>T_{c}$, $T=T_{c}$, $T_{K}<T<T_{c}$ and $T=T_{K}$. In these graphs the metastable part can be 
easily identified  by remembering that $W(q)$ must be a convex function of $q$.}
\label{figW}
\end{figure}

We introduce the Legendre transform $W(q)$ of the free energy $F(\eps)$:
\be
W(q)=F(\eps)+\eps q \ \ \ ; \ \ \ q={-\partial F \over \partial \eps} \ .
\ee
Analytic computation in mean field models \cite{pot}, as well as in  glass forming liquids 
using the replicated HNC 
approximation \cite{card}, show that the 
behaviour of $W(q)$ is qualitatively given by the  graphs of fig. \ref{figW}.

\begin{figure}
\includegraphics[width=0.45\textwidth]{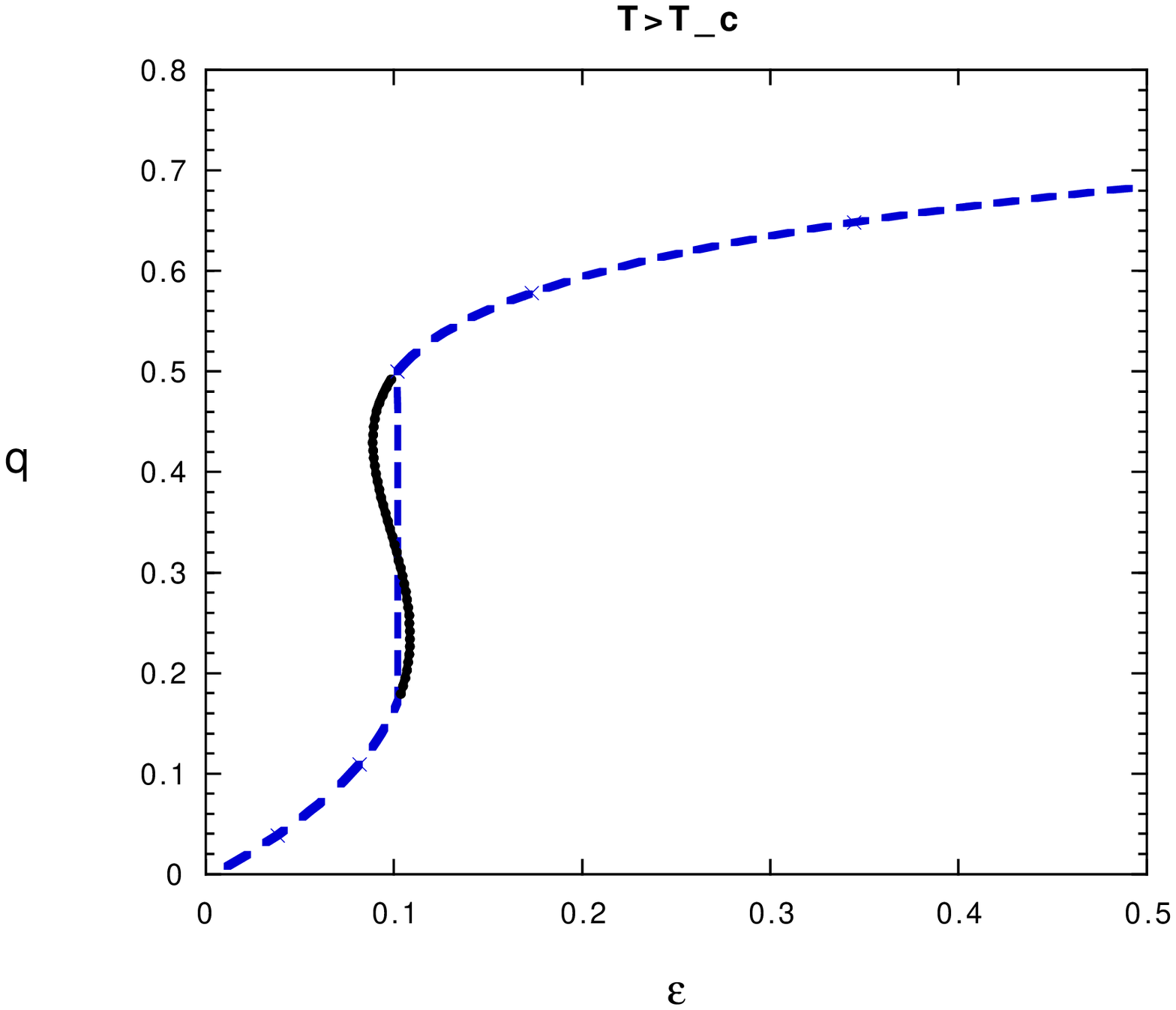}
\includegraphics[width=0.45\textwidth]{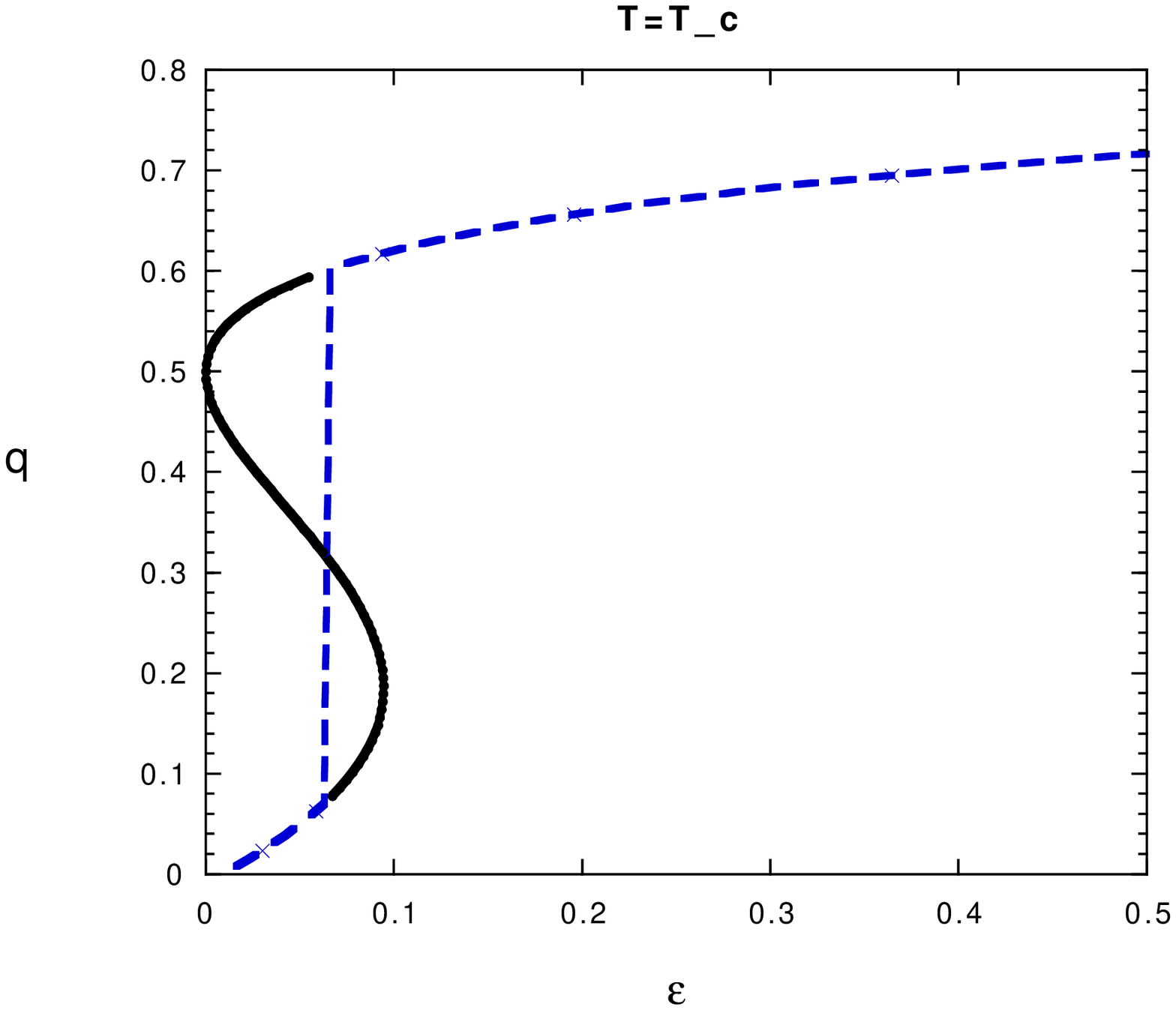}

\includegraphics[width=0.45\textwidth]{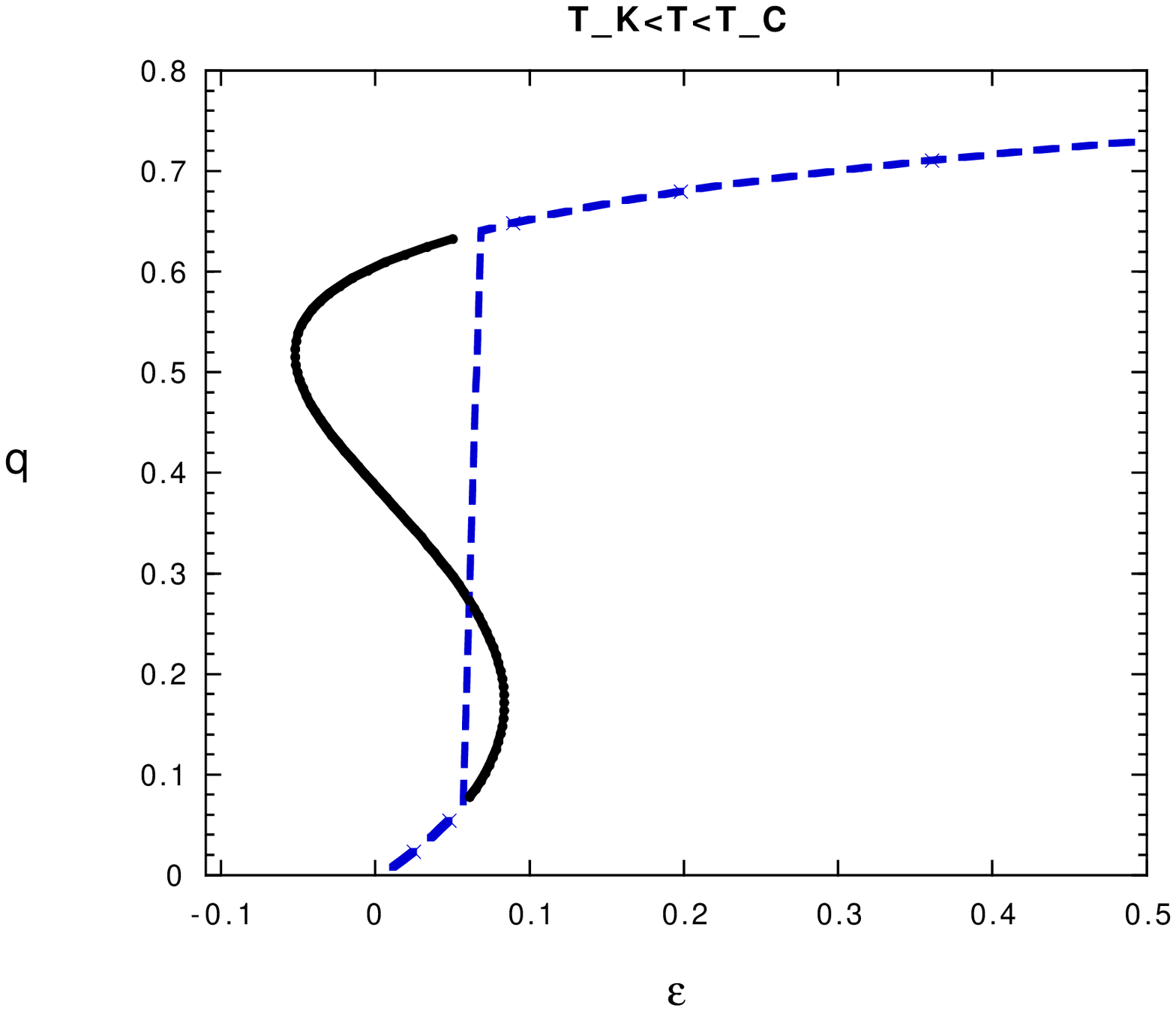}
\includegraphics[width=0.45\textwidth]{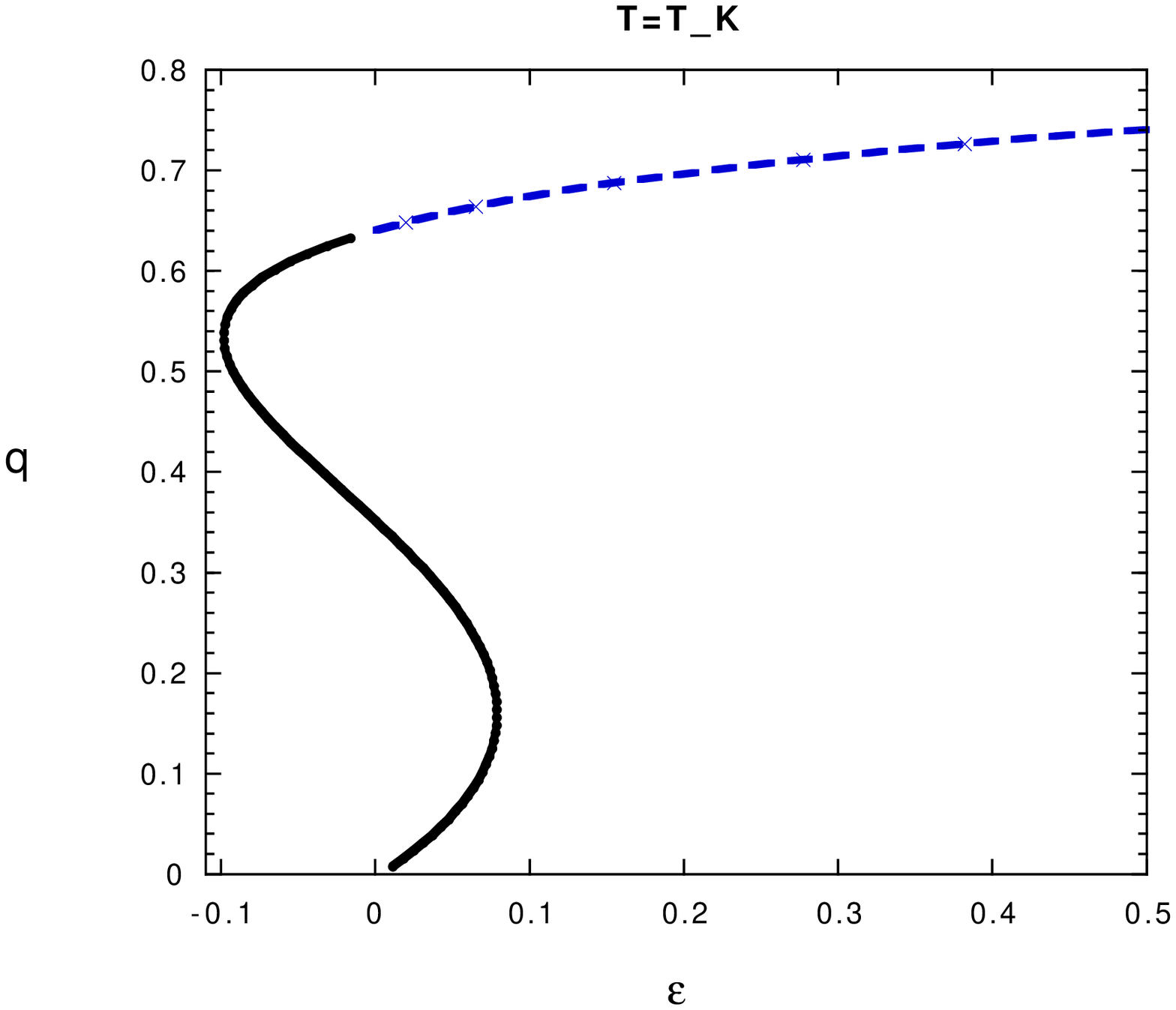}
\caption{Qualitative behaviour of the order parameter $q$, measuring the
typical  distance to
the reference configuration, versus the strength $\eps$ of the coupling
to this reference configuration,
 in the four regions
$T>T_{c}$, $T=T_{c}$, $T_{K}<T<T_{c}$ and $T=T_{K}$. The dashed line 
shows the true thermodynamically stable curve, where the full line 
is the metastable and the unstable part of the curve.
}
\label{figqeps}
\end{figure}

Fig. \ref{figqeps} shows the expectation  value of $q$ as function of $\eps$
in the corresponding four temperature ranges.
The results for the potential $W(q)$ in the unstable region where its second derivative is 
negative  and $q$ is a decreasing function of $\eps$
are a clear artefact of the mean field approximation, while the results in the metastable region  
correspond to phenomena that can be observed 
on time scales shorter than the lifetime of the metastable state.

The thermodynamic configurational entropy is the value of the potential
$W(q)$ at the secondary minimum with $q \ne 0$ \cite{pot}, and it
can be defined only if the minimum do exist (i.e. for $T<T_{c}$). It is evident that the secondary 
minimum for $T>T_{k}$ is always in the metastable region. 
 However if one would  start from a large value of $\eps$ 
and would  decrease $\eps$  to zero not too slowly,
the system would not escape from the metastable region and one 
obtains  a proper definition of the  thermodynamic configurational entropy in
this region $T>T_K$. In a similar way 
one could compute $q(\eps)$ in the region ($\eps > \eps_{c}$) where the high $q$ phase is 
thermodynamically stable
and extrapolate it to $\eps \to 0$.
The ambiguity in the definition of the thermodynamic configurational entropy 
at temperatures above $T_k$ becomes larger and 
larger when the temperature increases. It  cannot be defined for $T>T_{c}$.

\subsection{Numerical estimates of the configurational entropy}
Most attempts at estimating numerically the thermodynamic configurational
entropy start from the decomposition (\ref{Stdef}). The liquid entropy is estimated 
by a thermodynamic integration of the specific heat from the very dilute 
(ideal gas) limit. It turns out that in the deeply supercooled region
the temperature dependence of the liquid entropy is well fitted by the
law predicted in \cite{taraz}: $S_{liq}(T)=a T^{-2/5}+b$, which presumably allows
for a good extrapolation at temperatures $T$ which cannot be simulated.
As for the 'valley' entropy, it can be estimated as that of an harmonic
solid. One needs however the vibration frequencies of the solid. These have been 
approximated by several methods, which are all based on some
evaluation of the Instantaneous Normal Modes (INM) \cite{INM} in
the liquid phase,
and the
assumption that the spectrum of frequencies does not depend much on
temperature below $T_K$. Starting from a typical configuration
of the liquid, one 
can look at the INM  around it. In general there exist some negative eigenvalues
(the liquid is not a local minimum of the energy) which one must take care of. 
Several methods have been tried: either 
keep only the positive eigenvalues, or one considers the absolute values of the eigenvalues
\cite{sferesoft,LJ,LJ2}.
Alternatively one can also consider the INM around the nearest inherent structure
which has by definition a positive spectrum \cite{sferesoft,LJ,LJ2,SKT}. The
computation of the thermodynamic  entropy, using its definition as
a system  coupled to a reference
thermalized configuration, has also been studied in \cite{sferesoft}.

\begin{figure}
\includegraphics[width=0.4\textwidth,angle=270]{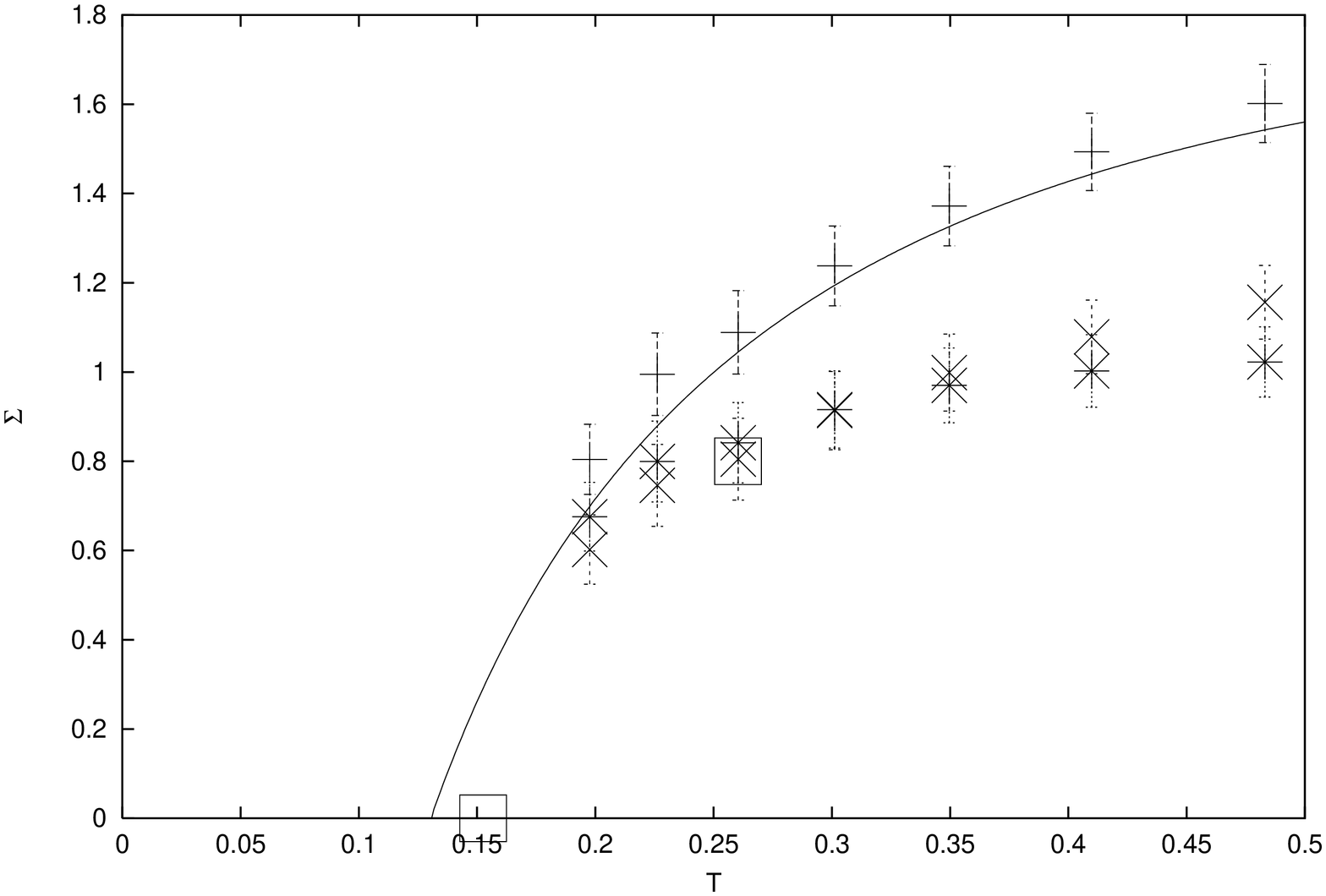}
\includegraphics[width=0.45\textwidth,angle=270]{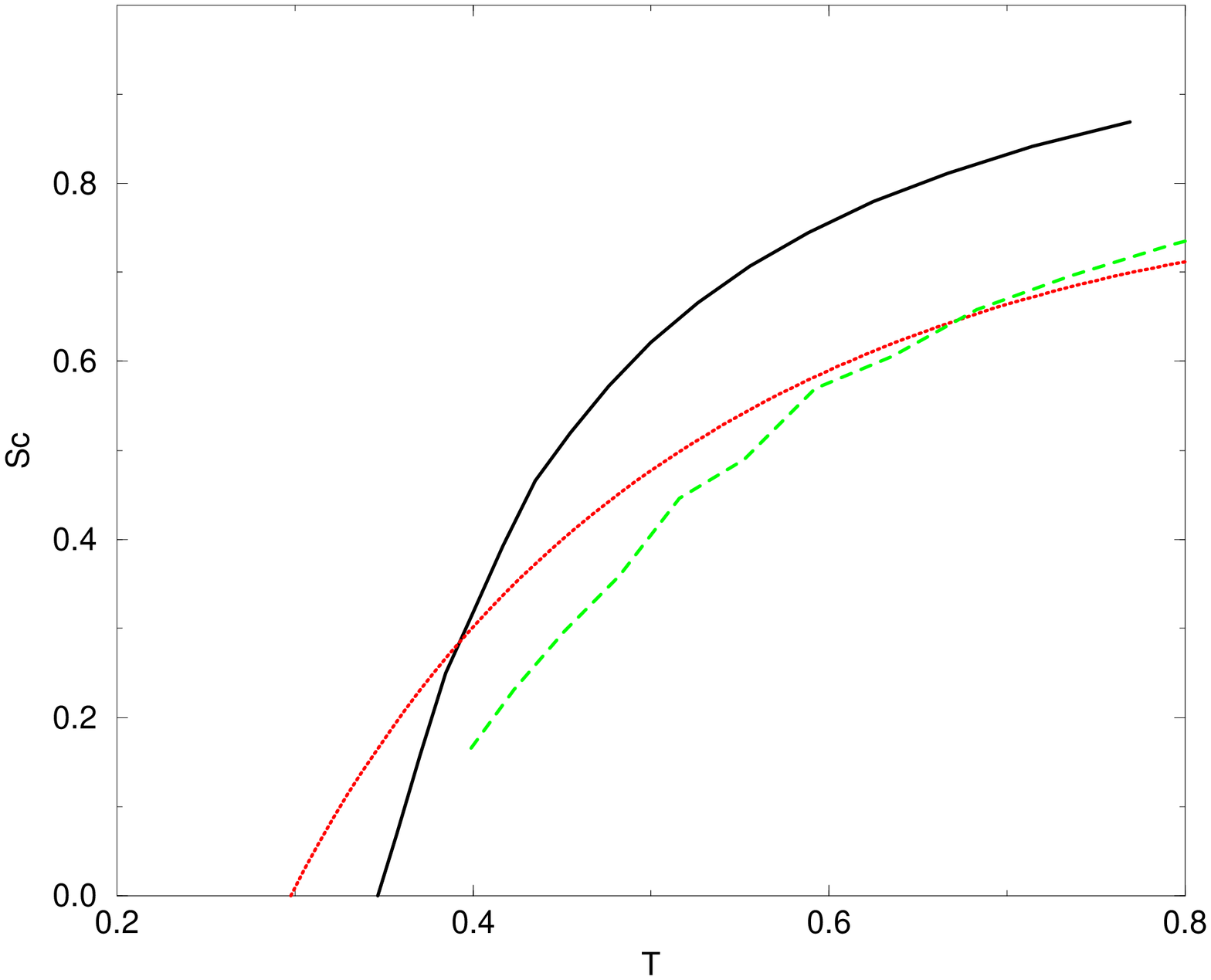}
\caption{The configurational entropy versus temperature in 
binary mixtures of soft-spheres and of   Lennard-Jones particles.
The soft sphere result (left curve), from \cite{sferesoft}, compares the analytical prediction
obtained within the harmonic resummation scheme (full line), to simulation
estimates of $S_{liq}-S_{valley}$, where the valley entropy is 
that of a harmonic solid with INM eigenvalues projected onto positive
eigenvalues (+), taken in absolute values ($\times$), or taken around
the nearest inherent structure ($\ast$). The  squares
correspond to
the numerical estimate of the
thermodynamic configurational entropy obtained by studying the system coupled  
to a reference configuration (see text, and \cite{sferesoft} for details). 
The Lennard-Jones result (right curve), shows as a 
full (black) curve the theoretical prediction obtained from the
cloned molecular liquid approach\cite{LJ,LJ2}. The dotted (green) curve is the result from
the simulations of \cite{LJ,LJ2} and the dashed (red) curve is the result
from the simulations of \cite{SKT}. Both simulations use the $S_{liq}-S_{valley}$
estimate where the harmonic solid vibration modes are approximated by the ones
of the nearest inherent structure.}
\label{figSc}
\end{figure}

The results for the configurational entropy as a function of temperature are
shown in fig. \ref{figSc}, for binary mixtures of soft spheres and of Lennard-Jones particles. 
The agreement with the analytical result obtained from the replicated fluid system
is rather satisfactory, considering the various approximations
involved both in the analytical estimate and in the numerical ones.

\begin{figure}
\includegraphics[width=0.65\textwidth,angle=270]{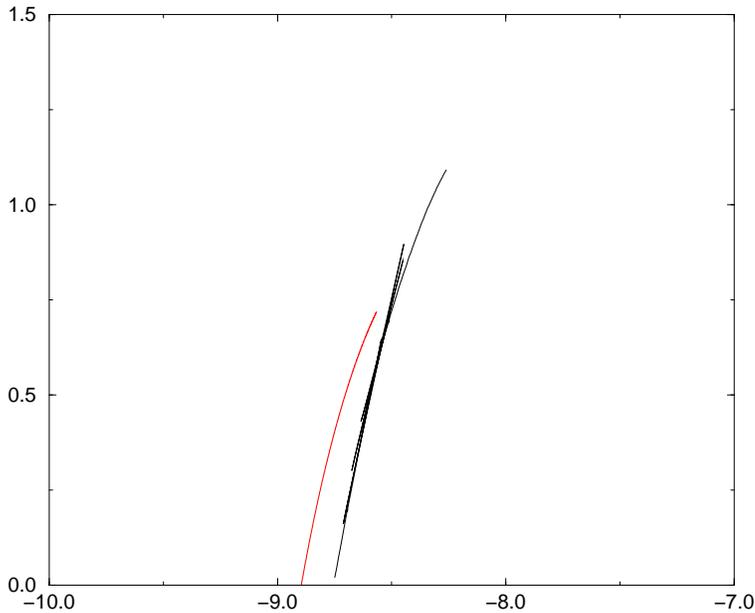}
\caption{
The left (red) curve is the configurational entropy of inherent structures
versus energy for a binary Lennard-Jones fluid, computed numerically
in \cite{SKT} (with respect to the curve plotted in \cite{SKT}, the
energies have been shifted in order to take into account the truncation of 
the Lennard-Jones potential used in the simulations of \cite{SKT}). The right
curve is the analytic prediction, using
the description of the molecular fluid of binary Lennard-Jones particles
of \cite{LJ,LJ2}. There is a small shift in energy between the two curves, 
but the overall agreement is satisfactory.
}
\label{scdee}
\end{figure}

In a recent work, Sciortino Kob and Tartaglia \cite{SKT} have computed the configurational
entropy of inherent structures, $\Si_{is}(T)$, defined in  (\ref{Sis}),
in binary Lennard-Jones system. 
Assuming that the free energy $-T \log Z(a)$ of an inherent structure $a$ 
($Z(a)$ is defined in (\ref{za}))
can be approximated by $E_a+\delta F(T)$, with a correction $\delta F$ which is
nearly independent of $E_a$, then the logarithm of the
probability of finding an inherent structure with a given energy $E_{IS}$ is
given by $-\beta E_{IS}+\Si_{is}(E_{IS}) +c^t$. One 
can thus deduce the $E_{IS}$ dependence of $\Si_{IS}$. Shifting the curves vertically
in order to try to superimpose them with the thermodynamic configurational entropy,
they have checked that all these curves coincide in the region of small enough
energy, confirming thus that these two definitions
of the configurational entropy agree at low enough energy or temperature.
In fig. \ref{scdee} we compare their result for the configurational entropy
of inherent structures to the one obtained analytically, using
the description of the molecular fluid of binary Lennard-Jones particles
of \cite{LJ,LJ2}. Apart from a small shift in the ground state energy
which may have several origins (finite size effects, small uncertainties in the 
description of the correlation in the molecular fluid),
the figures are in rather good agreement.

\section{Remarks}
We believe that we have now a consistent scheme for computing the 
thermodynamic properties of glasses at equilibrium. What is needed is on the one hand
some better approximations of the molecular liquid state, on the other hand some
precise numerical results in the glass phase at equilibrium, as 
well as measurements of the fluctuation dissipation ratio
in the  out of equilibrium dynamics (which should give the value of $m$ \cite{fdr}).
 Another
obvious direction is to study, with the present
methods,  various types of interaction potentials, including some which are characteristic
of strong glasses. Eventually, one would like to proceed to a first 
principle study of the out of equilibrium dynamics.
\section{Acknowledgments}
\label{acknowledgements}
We wish to thank W. Kob for providing the data discussed in the last section, and
for giving us the energy shift of the truncated Lennard-Jones problem, used
in the comparison of fig. \ref{scdee}. We wish to thank P. Verrocchio for
providing the analytic prediction shown in fig. \ref{scdee}.


\section{References}
\begin{thebibliography}{999}

\bi{glass_revue}
Recent reviews can be found in:  C.A. Angell, Science, 
{\bf 267}, 1924 (1995) and P.De Benedetti, `Metastable liquids', Princeton 
University
Press (1997). An introduction to the theory is: J.J\"ackle, Rep.Prog.
Phys. {\bf 49} (1986) 171.
\bi{kauzmann}
 A.W. Kauzman, Chem.Rev. {\bf 43} (1948) 219.
 A nice recent discussion can be found 
in R. Richert and C.A. Angell, J.Chem.Phys. {\bf 108} (1999) 9016.

\bi{AdGibbs}
G. Adams and J.H. Gibbs J.Chem.Phys {\bf 43} (1965) 139; J.H. Gibbs and E.A.
Di 
Marzio, 
J.Chem.Phys. {\bf 28} (1958) 373.

\bi{rubber}
In this sense, the rubber is a structural glass which is much closer to
spin glasses, because of the quenched random links between
the macromolecules. Theoretical studies of rubber
are reviewed in  P.M. Goldbart, H.E. Castillo and A. Zippelius
Adv. Phys. {\bf 45} (1996) 393.

\bi{MPV} For a review, see  M. M\'ezard, G. Parisi and M.A. Virasoro, {\sl Spin glass theory 
and 
beyond}, World Scientific (Singapore 1987)

\bi{REM}
B. Derrida, {\em Phys. Rev.}  {\bf B24}, 2613 (1981)

\bi{GrossMez}
D.J. Gross and M. M\'ezard, Nucl. Phys. {\bf B240} (1984) 431.


\bi{KiThWo}
 T.R. Kirkpatrick and P.G. Wolynes,  Phys. Rev. {\bf
A34}, 1045 (1986); T.R. Kirkpatrick and D. Thirumalai, Phys. Rev. Lett. {\bf 
58},
2091 (1987); T.R. Kirkpatrick and D. Thirumalai, Phys. Rev. {\bf B36}, 5388 
(1987); 
T.R. Kirkpatrick, D. Thirumalai and P.G. Wolynes,  Phys. Rev. {\bf
A40}, 1045 (1989).

\bi{crisanti} A. Crisanti, H. Horner and H.J. Sommers,
Z. Physik B {\bf 92}, 257 (1993).


\bi{nodis1}
 J.-P. Bouchaud and M. M\'ezard; J. Physique I (France) {\bf 
4} (1994) 1109.
E.  Marinari, G.  Parisi and F.  Ritort; J.  Phys.  {\bf A27} (1994) 7615; J.  
Phys.  {\bf A27} (1994) 7647.

\bi{nodis2}
P.Chandra, L.B.Ioffe and D.Sherrington, Phys. Rev. lett. {\bf 75} (1995) 713,
and cond-mat/9809417.
P.Chandra, M.V. Feigelman and L.B.Ioffe, Phys. Rev. lett. {\bf 76} (1996) 4805.

\bi{nodis3}
 E. Marinari, G. Parisi and F. Ritort, cond-mat/9410089.
 S. Franz and  J. Hertz, {\it Phys. Rev. Lett.} {\bf 74}, 2114 (1995).

\bi{cuku} L.  F.  Cugliandolo and J.Kurchan, Phys.  Rev.  Lett.  {\bf 71}, 
1 (1993).

\bibitem{fdr} S. Franz, M. M\'ezard, G. Parisi and L. Peliti,
Phys. Rev. Lett. {\bf 81} 1758 (1998); {\em The response of glassy systems
 to random perturbations: A bridge between equilibrium and 
off-equilibrium}, cond-mat/9903370, to appear in J.Stat.Phys.

\bi{gpglass}
 G. Parisi Phys.Rev.Lett. {\bf 78}(1997)4581.

\bi{bk1}
 W. Kob and J.-L. Barrat, Phys.Rev.Lett. {\bf 79} (1997) 3660.

\bi{bk2}
 J.-L. Barrat and W. Kob, cond-mat/9806027.

\bi{MePa1} M. M\'ezard and G. Parisi, Phys.  Rev.  Lett.  {\bf 82}, 747 
(1998).
 
\bi{Me} M. M\'ezard, Physica A {\bf 265}, 352 (1999).

\bi{MePa2} M. M\'ezard and G. Parisi J. Chem.  Phys. {\bf 111}, 1076 (1999).

\bi{sferesoft} B. Coluzzi, M. M\'ezard, G. Parisi and P. Verrocchio, {\em
Thermodynamics of binary 
mixture glasses}, cond-mat/9903129.

\bi{LJ} B. Coluzzi, G. Parisi and P. Verrocchio, {\em Lennard-Jones
  binary mixture: a 
thermodynamical approach to glass transition}, cond-mat/9904124.

\bi{LJ2} B. Coluzzi, G. Parisi and P. Verrocchio, {\em The thermodynamical
liquid-glass transition in a Lennard-Jones binary mixture}, 
cond-mat/9906124.

\bi{kepler} An introduction to recent work on  Kepler's conjecture can be
found in:
www.math.lsa.umich.edu/~hales/countdown/.

\bi{angelani}
 L. Angelani, G. Parisi,
G. Ruocco and G. Viliani, cond-mat/9904125.

\bi{theo}
T.M. Nieuwenhuizen, Phys.Rev.Lett. {\bf 79} (1997) 1317.

\bi{inherent}
M. Goldstein, J. Chem. Phys. {\bf 51}, 3728 (1969);
 F.H. Stillinger, Science {\bf 267} (1995) 1935, and references 
therein. Recent includes: S. Sastry, P.G. Debenedetti and
F.H. Stillinger, Nature {\bf 393}, 554 (1998),
W. Kob, F. Sciortino and P. Tartaglia, cond-mat/9905090;
 F. Sciortino, W. Kob  and P. Tartaglia, cond-mat/9906278;
S. B\"uchner and A. Heuer, cond-mat/9906280.


\bi{remi} R. Monasson, {\em Phys. Rev. Lett.} {\bf 75}, 2847 (1995).


\bi{MEPAZEE} M. M\'ezard, G. Parisi and A. Zee {\em Spectra of Euclidean 
Random
Matrices}, 
cond-mat/9906135.

\bi{CAGIAPA} A. Cavagna, I. Giardina and G. Parisi, {\em Analytic 
computation of
the Instantaneous 
Normal Modes spectrum in low density liquids} (cond-mat/9903155),
 Phys.Rev.Lett.  to be 
published.

\bi{taraz}
Y.Rosenfeld and P. Tarazona, Mol.Phys. {\bf 95}, 141 (1998).

\bi{pedago}
G.Parisi,  cond-mat/9905318.


\bi{Han2}
B. Bernu, J.-P. Hansen, Y. Hitawari and G. Pastore, {\em Phys. Rev.} 
{\bf A 36}, 4891 (1987). J.-L. Barrat, J.-N. Roux and J.-P. Hansen, 
{\em Chem. Phys.} {\bf 149}, 197 (1990). J.-P. Hansen and S. Yip, 
{\em Trans. Theory and Stat. Phys.} {\bf 24}, 1149 (1995).


\bi{SPEEDY} See the talk of Speedy at this conference and references therein. 

\bi{pot}  S. Franz ang G. Parisi, J. Physique I {\bf 5} (1995) 1401; 
 Phys.Rev.Lett.  {\bf 79} (1997) 2486.


\bi{card} 
 M.Cardenas, S. Franz and G. Parisi, cond-mat/9712099.

\bi{INM} T. Keyes, J. Chem. Phys. A101 (1997) 2921.

\bi{SKT} F. Sciortino, W. Kob and P. Tartaglia, { \em Inherent structure entropy
of supercooled liquids}, cond-mat/9906081. See also Sciortino's contribution
to this volume.


\end{thebibliography}
\end{document}